\documentclass[prl,aps,preprint,groupedaddress,floats,showpacs]{revtex4}

\usepackage{graphicx}
\usepackage{makeidx,epsfig}

\usepackage{amssymb}

\begin{document}

\title{Worm Algorithm for Problems of Quantum and Classical Statistics}

\author{Nikolay Prokof'ev and Boris Svistunov}

\address{Department of Physics, University of
Massachusetts, Amherst, MA 01003, USA}

\begin{abstract}
 This is a chapter of the multi-author book ``Understanding Quantum Phase
Transitions," edited by Lincoln Carr and published by Taylor \& Francis. In
this chapter, we give a general introduction to the worm algorithm and present
important results highlighting the power of the approach.
\end{abstract}

\maketitle

Theoretical studies often involve mappings of the original system
onto an equivalent (with regards to the final answer for some property)
description in terms of abstract mathematical/graphical objects. Path integrals,
high-temperature expansions, and Feynman diagrams are the well-known examples considered in
this chapter. Under mapping, one has to deal with the infinite-dimensional configuration
space having complex topology and non-local constraints which severely reduce efficiency
of Monte Carlo simulations based on standard local updates.
This sometimes leads to ergodicity problems in large system when the entire configuration
space can not be sampled in a reasonable computation time.
A somewhat related difficulty facing conventional Monte Carlo schemes is the
computation of off-diagonal correlation functions since they have no
direct relation to the configuration space of the partition function.
In what follows we consider path integrals for lattice and continuous systems,
high-temperature expansions, and  Feynman diagrams, explain the general idea of how
Worm Algorithms deal with the topological constraints by going
to the enlarged configuration space, and present illustrative results for
several physics problems.

\section{Path-Integrals in Discrete and Continuous Space}
\label{PS:sec:paths}

For clarity, we start by introducing the Hamiltonian describing lattice bosons
(a straightforward generalization to fermions will be mentioned later)
making hopping transitions between the nearest-neighbor sites $<ij>$,
interacting by the pairwise potential $U_{ik}$, and subject to the external potential
$\mu_i$:
\begin{equation}
H = H^{(0)} + H' = \frac{1}{2} \sum_{i,k} U_{ik} \: n_{i} n_{k} -
\sum_{i} \mu_{i}  n_{i}  - t \sum_{<ij>} b_{j}^{\dag} b_{i}^{\:} \;.
\label{PS:BHmodel}
\end{equation}
Here $b_{i}$ is the bosonic annihilation operator and $n_{i}=b_{i}^{\dag}b_{i}$.
In the Fock basis of site occupation numbers, $\vert  \alpha \rangle  =\vert  \{ n_i \} \rangle$,
the first two terms, representing $H^{(0)}$,
are diagonal, while the last term, representing $H'$, is not. We write the
statistical operator as
\begin{equation}
e^{-\beta H } = e^{-\beta H^{(0)} } \exp \left\{ -\int_0^{\beta} d\tau H'(\tau )  \right\}   \;,
\label{PS:SO1}
\end{equation}
where $H'(\tau ) = e^{\tau H^{(0)} } H' e^{-\tau H^{(0)} }$, and the exponential is understood as
the time-ordered expansion
\begin{equation}
1 - \int_0^{\beta}d\tau \: H'(\tau )  + \int_0^{\beta} d\tau_1   \int_{\tau_1}^{\beta}  d\tau_2 \: H'(\tau_1 ) H'(\tau_2 )  +  \dots \;.
\label{PS:SO2}
\end{equation}

\begin{figure}
\resizebox{\hsize}{!}{
\includegraphics[width=120pt, height=150pt, angle=-90]{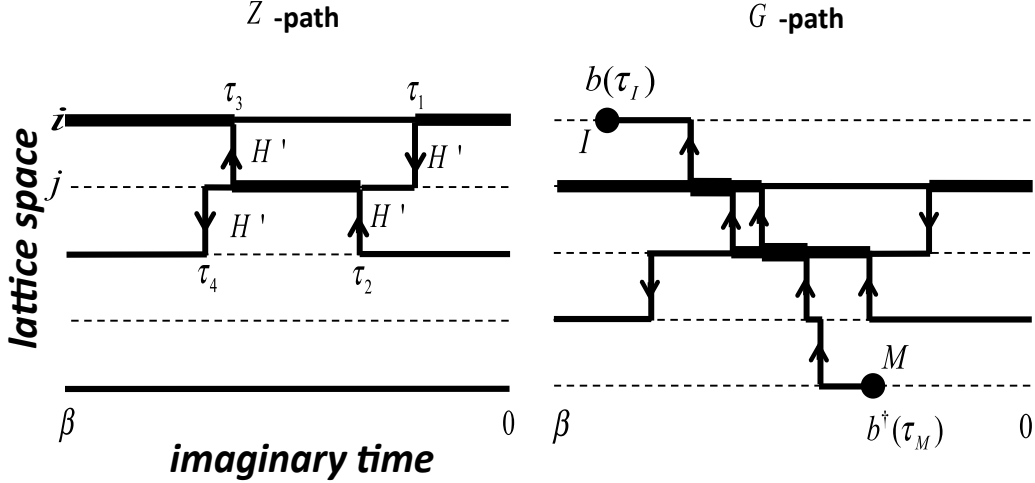} }
\vspace*{-3.5cm}
\caption{Lattice path-integral representations for the partition function and the
Green's function. Line thickness is proportional to $n_i$}
\label{PS:paths}
\end{figure}

Lattice path-integrals immediately follow from the graphical representation of the expansion (\ref{PS:SO2}).
Consider the partition function $Z$ given by the trace of $e^{-\beta H }$ and write explicitly
the $m$-th term (before integration over time) as
\begin{equation}
(-1)^m\,  d^m \tau  \, \ e^{-(\beta-\tau_1)H^{(0)}_{\alpha_0}} \left( H'_{\alpha_0,\alpha_1} \right) \: e^{-(\tau_1-\tau_2)H^{(0)}_{\alpha_1}} \dots \: \left(
 H'_{\alpha_{m-1},\alpha_m}  \right) \:
 e^{-\tau_m H^{(0)}_{\alpha_m}} \;,
\label{PS:SOnth}
\end{equation}
with $\alpha_m \equiv \alpha_0 $ to reflect the trace condition
(periodic boundary condition in imaginary time), and $d^m \tau \equiv d\tau_1 \dots d\tau_m$.
Since hopping terms in $H'$ change the state by shifting only one particle to a nearest-neighbor site,
the sequence of matrix elements is completely determined by specifying the
``evolution'' or ``imaginary time trajectory'' of occupation numbers $\{ n_i (\tau ) \}$.
In the left panel of Fig.~\ref{PS:paths} we show the trajectory describing one of the
$4$-th order terms which contributes
$ t^4\,  d^4 \tau \, 1 \cdot 2 \cdot \sqrt{2} \cdot \sqrt{2} \: \exp \{-\int_0^{\beta } 
d\tau H^{(0)}(\tau) \} $
to $Z$, where for brevity we use $H^{(0)}(\tau)$ for the energy of the state
$|\{ n_{i}(\tau) \} \rangle $ and give explicit expressions for the hopping matrix elements
$\langle n_{i}-1, n_{j}+1 | -t b_{j}^{\dag} b_{i}^{\:}|n_{i}, n_{j} \rangle =
-t \sqrt{n_{i} (n_{j}+1)}$.
Thus, the partition function can be written as a sum over all possible paths
$\{ n_{i} (\tau ) \}$ such that $n_{i} (\beta ) = n_{i} (0 )$
\begin{equation}
Z = \sum_{\{ n_{i} (\tau ) \} } W[\{ n_{i} (\tau ) \} ]  \;,
\label{PS:LPI}
\end{equation}
with strict rules relating the trajectory shape to its contribution to $Z$.
The trajectory weight is sign-positive if $t$ is positive or the lattice is bi-partite.

The path-integral language at this point is nothing but a convenient way of visualizing
each term in the perturbative expansion (\ref{PS:SO2}).
Due to the particle number conservation, the many body trajectory can be decomposed into
the set of closed (in the time direction) single-particle trajectories, or worldlines.
Worldlines can ``wind" around the $\beta$-circle several times before closing on
themselves. In a system with periodic boundary conditions in space, the trajectories can
also wind in the space direction.
Worldlines with non-zero $\beta$- and space-winding numbers are said to form exchange cycles
and winding numbers, respectively; they are directly responsible for superfluid properties
of the system \cite{PS:Pollock} and are the origin of non-local topological constraints mentioned
in the introductory paragraph.

The Green's function of the system,
\begin{equation}
G(\tau_M-\tau_I , i_M-i_I) = T_{\tau} \langle b_{i_M}^{\dag} (\tau_M ) b_{i_I}^{\:} (\tau_I) \rangle \;,
\label{PS:Gfunction}
\end{equation}
has a similar path-integral representation, see right panel in Fig.~\ref{PS:paths},
with one notable difference: due to
operators $b^{\dag}$ at point $(i_M,\tau_M)$ and $b^{\:}$ at point $(i_I,\tau_I)$
there is one more particle present in the system on
the time interval $(\tau_I, \: \tau_M)$, i. e.
when the $\{ n_{i} (\tau ) \}$ evolution is decomposed into worldlines there
will be one {\it open} worldline originating at point $\tau_M$ and terminating
at point $\tau_I$. These special points will be labeled as $\cal I$ (``{\it Ira}'') and $\cal M$
(``{\it Masha}'') throughout the text. We will use short-hand notations, $Z$-path,
and $G$-path to distinguish configuration spaces of $Z$ and $G$.

We skip here the standard derivation of Feynman's path-integral representation
for an interacting $N$-particle Hamiltonian in continuous space \cite{PS:RMP48}
\begin{equation}
H = K+V = - \frac{1}{2m} \sum_{i=1}^{N} \nabla_i^2 + \frac{1}{2} \sum_{i \ne j =1}^{N} U(r_i-r_j) - \sum_{i=1}^{N} \mu(r_i) \;.
\label{PS:Hcont}
\end{equation}
leading to
\begin{equation}
Z=  \oint {\cal D} {\mathbf R}_\tau  \
\exp \left\{  - \int_0^\beta \left[ \frac{ m \dot{\mathbf R}^2}{2} +  U({\mathbf R}) -\mu ({\mathbf R})
\right]  d\tau  \right\}  \;.
\label{PS:Zcont4}
\end{equation}
where $R_\tau  = \{ r_i(\tau ) \} $ represents positions of all particles at time $\tau$,
$U(R)$ and $ -\mu (R)$ stand for internal and external potential energy respectively,
and $R_{\beta} \equiv R_{0}$ to satisfy the trace condition. In practice,
the imaginary time axis is sliced into sufficiently large number of intervals and the trajectory
is defined by specifying particle positions at a discreet set of time points.
Apart from time slicing, there is no fundamental difference between the lattice
and continuous path-integrals. The $Z$-path consists of closed worldlines
(the decomposition of the many-body path into individual worldlines is unique in continuous space),
and the $G$-path contains one {\it open} worldline originating at $(r_M, \tau_M)$ and terminating at
$(r_I, \tau_I)$.

The only difference between the fermioninc and bosonic systems is in the sign rule:
for fermions the trajectory weight $W$ involves an additional factor $(-1)^p$,
where $p$ is the parity of the permutation between the particle coordinates in $R_{\beta}$
relative to $R_{0}$.

\section{Loop Representations for Classical High-Temperature Expansions}
\label{PS:sec:loop_HTE}

\begin{figure}
\resizebox{\hsize}{!}{
\includegraphics[width=80pt, height=100pt, angle=-90]{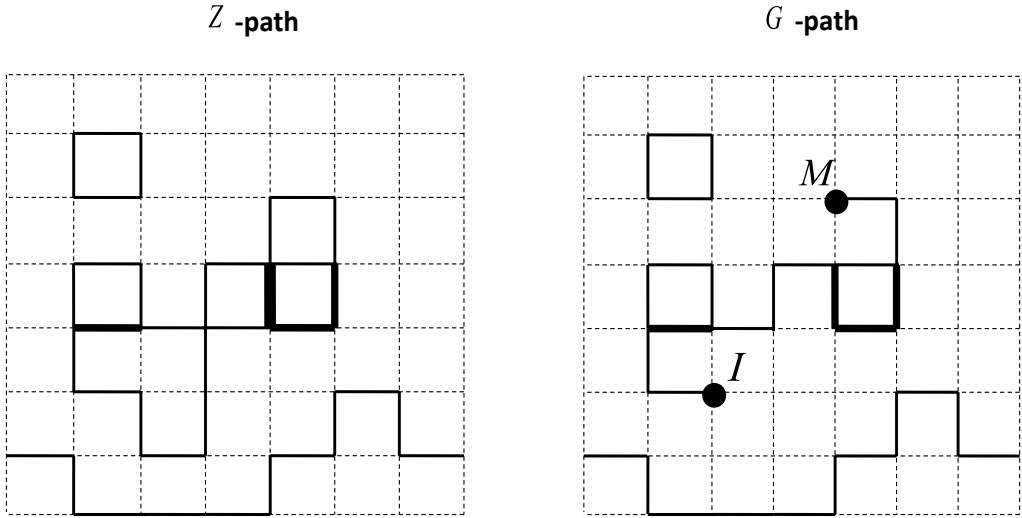} }
\vspace*{-2.5cm}
\caption{Loop representations for the partition function and the
correlation function for the Ising model. Line thickness is proportional to $N_b$.}
\label{PS:fig:loops}
\end{figure}

Classical statistical models can be also mapped to the configuration space of closed loops.
Important examples include Ising, XY, and  $\vert \psi \vert^4$ models, as well as
their multi-component generalizations with or without the gauge coupling, e.g. the $CP^1$ model.
We illustrate the general idea of the mapping by considering the simplest case of the
Ising model when $-H/T= K \sum_{<ij>} \sigma_i \sigma_j$ for $N$ Ising spin variables $\sigma_i=\pm 1$.
One starts with the partition function and expands bond Gibbs factors into Taylor series
(to simplify notations we will use subscript $b$ for lattice bonds)
\begin{equation}
Z=\sum_{\{ \sigma_i=\pm 1 \}} \prod_{b=<ij>} e^{K \sigma_i \sigma_j} =
\sum_{\{ \sigma_i=\pm 1 \}} \prod_{b=<ij>} \sum_{N_b=0}^{\infty}
\frac{K^{N_b}}{N_b!} \left( \sigma_i \sigma_j \right)^{N_b} \;.
\label{PS:I1}
\end{equation}
By changing summation over $\{ N_b \}$ and $\{ \sigma_i=\pm 1 \}$ places we obtain
\begin{equation}
Z=2^N \sum_{\{ N_b \}}^{loops} \prod_{b=<ij>} \frac{K^{N_b}}{N_b!} \equiv
2^N \sum_{\{ N_b \}}^{loops} W[\{ N_b \} ]  \;.
\label{PS:I2}
\end{equation}
The ``loops'' label on the sum represents the constraint that the sum of all bond numbers
incident on every lattice site, $L_i=\sum_{b=<ij>} N_b$, has to be even; otherwise $\sum_{ \sigma_i \pm 1} \sigma^{L_i}$ is zero. 
In the graphical representation where $N_b$ is substituted with $N_b$ lines
drawn on the corresponding bond, this is equivalent to demanding that the allowed configuration
of lines is that of closed un-oriented loops, since loops always contribute an 
even number to $L_i=\sum_{b=<ij>} N_b$, see the left panel in Fig.~\ref{PS:fig:loops}.  
[Using identity $e^{K \sigma_i \sigma_j} = \cosh(K) \sum_{N_b=0,1}
\left[ \tanh(K) \sigma_i \sigma_j  \right]^{N_b}$ we get a more compact
formulation since only one line can be drawn on the bond.]
In close analogy with the
Green's function, the configuration space of the spin-spin correlation function
$G_{IM}=\langle \sigma_I \sigma_M \rangle$ is that of closed loops with one open line
originating at site $\cal I$ and terminating at $\cal M$, see the right panel in Fig.~\ref{PS:fig:loops}.
The difference between the XY and Ising models is that loops are oriented
in the XY-case, i.e. $N_b \in (-\infty, \infty )$.

\section{Worm Algorithm: The Concept and  Realizations}
\label{PS:sec:WA}

The Worm Algorithm (WA) strategy for updating loop configuration spaces
involves two major ideas: \\
\noindent {\bf 1.} The configuration space is enlarged to include one
open line, as if someone started drawing a new loop but is
not finished yet. In all examples mentioned above this is not merely
an algorithmic trick but also an important tool to have direct access to
off-diagonal correlation functions, such as the Green's function \cite{PS:WA1, PS:WA2} (with two worms
one may calculate multi-particle off-diagonal correlations as well \cite{PS:SF_SF,PS:Soyler}.)
From time to time the two ends of the open line come close
and get connected thus making a loop and transforming $G$-path to $Z$-path.
In Monte Carlo, drawing and erasing processes are balanced and are
complementary to each other. \\
{\bf 2.} All updates on $G$-paths are performed {\it exclusively}
through the end-points of the open line, no global updates or local updates
transforming one $Z$-path to another $Z$-path are necessary.

More generally, the Worm Algorithm idea is to consider an enlarged configuration space
which includes structures violating constraints present in the $Z$-sector
of the space. Often, this can be achieved automatically by considering the
relevant correlation functions. Green's function and, for paired states,
higher-order off-diagonal correlators are the appropriate choice for the
path-integral space. However, one should feel free to introduce ``unphysical''
configurations which bear no meaning at all and are used solely
for employing local updates to produce non-trivial global changes of physical configurations.
An example with the momentum conservation law in Feynman diagrams discussed
below illustrates the point.

WA is a local Metropolis scheme, but, remarkably, its efficiency is similar to 
(or better than) the best cluster methods at the critical point, i.e. it does not suffer from the critical
slowing down problem. It has no problem producing loops winding around the system,
allows efficient simulations of off-diagonal correlations, grand canonical ensembles,
disordered systems, etc. Below we provide more specific details of how it works.

\begin{figure}
\resizebox{\hsize}{!}{
\includegraphics[width=120pt, height=150pt, angle=-90]{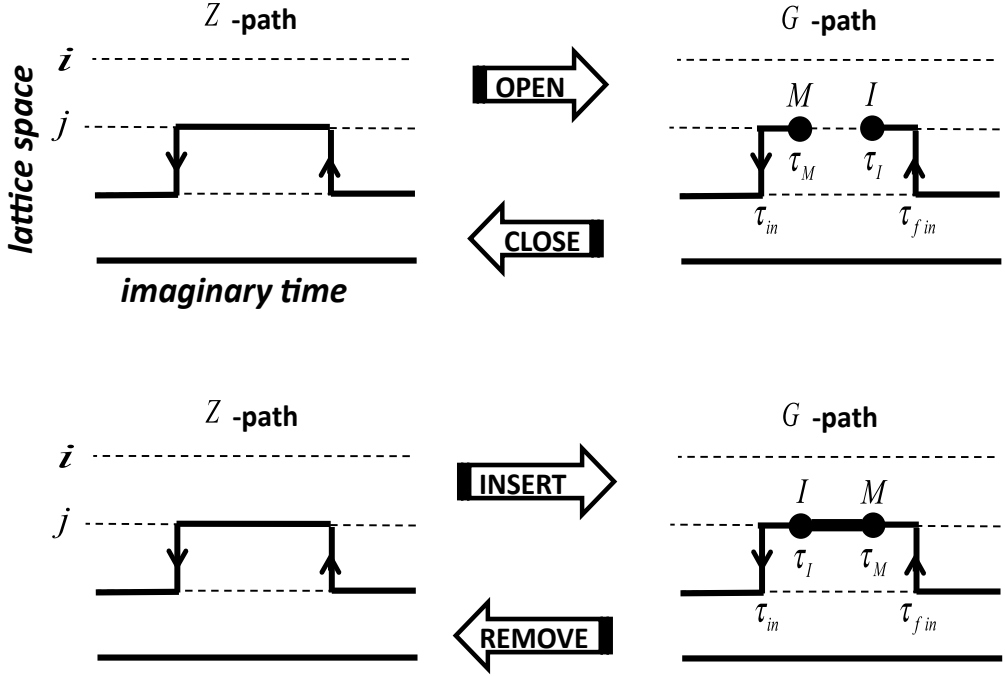} }
\vspace*{-1.0cm}
\caption{ The upper and lower panels illustrate transformations performed by the \textit{Open/Close}
and  \textit{Insert/Remove} pairs of updates respectively on the lattice path-integral configurations.}
\label{PS:fig:open_close}
\end{figure}

\subsection{Discrete Configuration Space: Classical High-\-Tempe\-rature Expansions}
\label{PS:subsec:HTE}

We start with the simplest case of the classical Ising model. The entire algorithm
consists of just one update: \\
If ${\cal I}={\cal M}$,  select at random a new lattice site $j$ and assign ${\cal I} ={\cal M}=j$;
otherwise skip this step. [In other words, put your pensil/eraser anywhere.]
Select at random the direction (bond) to shift {\it Masha} to a n.n. site, let it be $k$, and
propose to change the bond number from $N_b$ to $N_b'={\rm mod}_2(N_b+1)$.
Accept the move with probability $R=\max[1,\tanh^{N_b'-N_b}(K)]$.
This is a complete description of the algorithm!

Every configuration contributes unity to the statistics of $G_{I,M}$.
For the Ising model $Z=G_{I=M}$. Generalizations to other classical statistical models
are straightforward \cite{PS:WA1}. Ergodicity is guaranteed because a finite
number of steps is required to erase any initial trajectory and then to draw a
new one, line after line. The efficiency is ultimately linked to the fact
that WA works directly with the correlation function of the order parameter field.

\subsection{Continuous Time: Quantum Lattice Systems}
\label{PS:subsec:cont_time}

We now illustrate the WA updating strategy for the system of lattice bosons
described by Eqs.~(\ref{PS:BHmodel}), (\ref{PS:LPI}) with the configuration space
shown in Fig.~\ref{PS:paths}. The set of updates presented below
forms an ergodic set. It is sufficient to describe updates performed with the {\it Masha}-end
of the open worldline; updates involving the {\it Ira}-end follow immediately
from the time reversal symmetry.

\begin{figure}
\resizebox{\hsize}{!}{
\includegraphics[width=110pt, height=150pt, angle=-90]{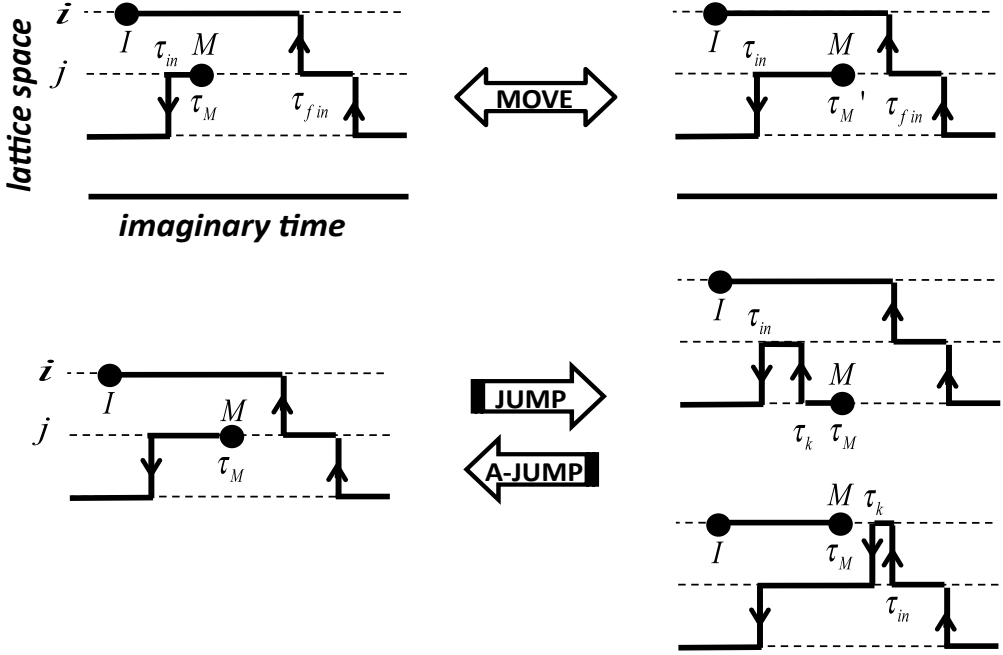} }
\vspace*{-1.2cm}
\caption{The upper and lower panels illustrate transformations performed by the \textit{Move}
and  \textit{Jump/anti-Jump} updates respectively on the lattice path-integral configurations.}
\label{PS:fig:move_jump}
\end{figure}

{\bf {\it  Open/Close}}.---This pair of updates takes us back and forth between the
$Z$- and $G$-paths by selecting an existing worldline and erasing
a small part of it (\textit{open}) or drawing a small piece of worldline between the end-points
of the open line to complete the loop (\textit{close}). These updates are illustrated in the upper
panel of Fig.~\ref{PS:fig:open_close}. In the (\textit{open}) update the path interval characterized by
time-independent occupation numbers on a given site is selected at random from the list of such intervals,
and the imaginary times $\tau_I < \tau_M$ are seeded from the normalized probability distribution
$p(\tau_I, \tau_M)$ with $\tau_I, \tau_M \in (\tau_{\rm in}, \tau_{\rm fin})$. Formally, the distribution
function is arbitrary, and this freedom should be used to optimize the acceptance ratio
$R \propto W_{\rm new}/[W_{\rm old}p(\tau_I, \tau_M)]$.
In the (\textit{close}) update one checks whether {\it Ira} and {\it Masha} are connected by a single
path interval with time-independent occupation numbers on a given site, and if true,
proposes to eliminate them by ``connecting the dots''. We skip here further technical details
(which are minimal), as well as explicit expressions for acceptance ratios, which can be found
elsewhere \cite{PS:WA2}.

{\bf {\it Insert/Remove} }.---This pair of updates also switches back and forth between the
$Z$- and $G$-paths by drawing a small piece of a new worldline (\textit{insert})
or erasing a small piece of worldline between the end points (\textit{remove}). The two updates are
illustrated in Fig.~\ref{PS:fig:open_close} and in all respects are similar to the {\textit{Open/Close} pair
except that the time ordering of $\tau_I$ and $\tau_M$ is reversed, see the lower panel in Fig.~\ref{PS:fig:open_close}.

{\bf  {\it Move}}.---Once in the $G$-path space the algorithm is
proposing to move {\it Masha} along the time axis, $\tau_M \to \tau_M'$ within the bounds
determined by the change of the occupation numbers on a given site. This motion
is equivalent to drawing/erasing the worldline, see the upper panel in Fig.~\ref{PS:fig:move_jump}.

{\bf  {\it Jump/anti-Jump}}.---This is the only complementary pair of updates
which is different from continuous transformations of lines since it involves the motion
of end-points in space. Without changing the time position of {\it Masha}, we place it on the neighboring site
and connect worldlines of the two sites involved in the update in such a way that
the rest of the path remains intact. This requires adding/removing a kink immediately before or after
$\tau_M$. We illustrate the \textit{Jump/anti-Jump} updates in the lower panel of Fig.~\ref{PS:fig:move_jump}.
Note the difference between the two cases. Whenthe kink is inserted to the left of {\it Masha}, 
the transformation can still be interpreted as proceeding with drawing the same worldline. 
When the kink is inserted to the right of {\it Masha}, we reconnect existing worldlines and ultimately 
effectively sample all allowed topologies of the path.

\begin{figure}
\resizebox{\hsize}{!}{
\includegraphics[width=110pt, height=150pt, angle=-90]{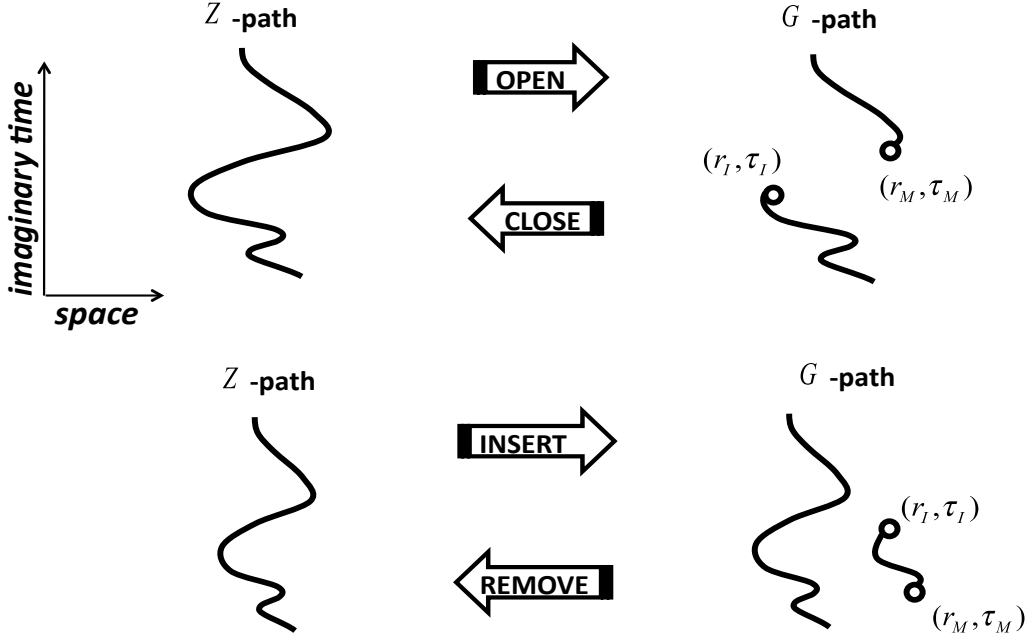} }
\vspace*{-1.5cm}
\caption{ The upper and lower panels illustrate transformations performed by the \textit{Open/Close}
and  \textit{Insert/Remove} pairs of updates respectively on the continuous-space path-integral configurations.}
\label{PS:fig:cont_space}
\end{figure}

This concludes the description of the algorithm. Such properties as density, energy, 
density-density correlations, etc. are computed using standard rules when the configuration is in the $Z$-path sector.
Every $G$-configuration makes a direct contribution to the statistics of
$G(\tau_M-\tau_I , i_M-i_I)$.

\begin{figure}
\resizebox{\hsize}{!}{
\includegraphics[width=110pt, height=150pt, angle=-90]{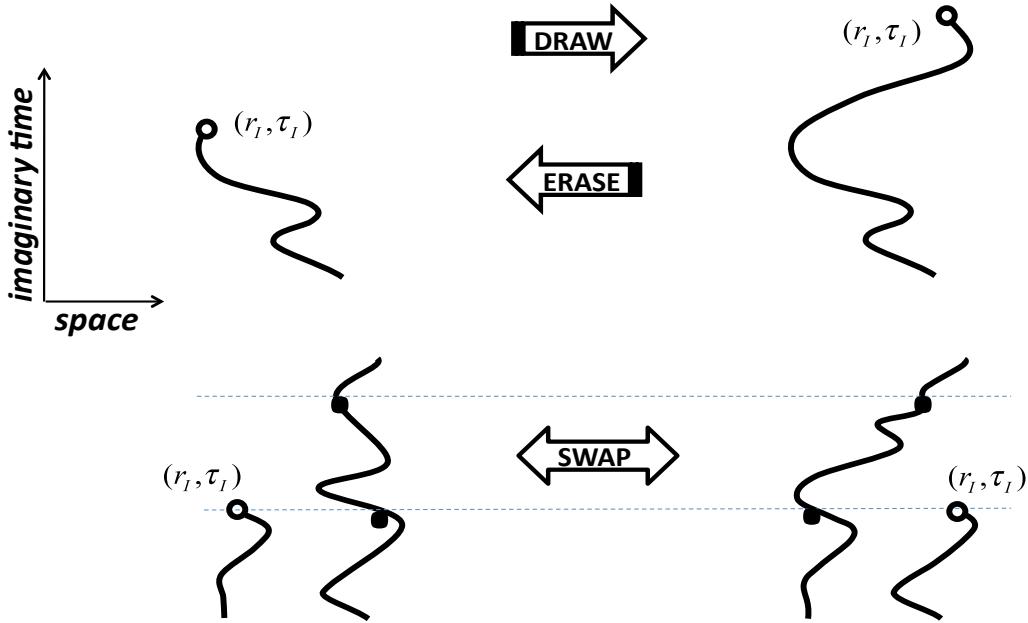} }
\vspace*{-1.5cm}
\caption{ The upper and lower panels illustrate transformations performed by the \textit{Draw/Erase}
and  \textit{Swap} updates respectively on the continuous-space path-integral configurations.}
\label{PS:fig:swap}
\end{figure}

\subsection{Bosons in Continuous Space}
\label{PS:subsec:cont_space}

Worm Algorithms for the continuous and lattice systems are essentially identical
at the conceptual level. Most differences are technical and originate from having discreet
instead of continuous time and continuous instead of discrete space. Specific protocols of
how one proposes new variables and performs measurements can be found in Ref.~\cite{PS:PRE}.
The only difference worth mentioning is that in continuous space the decomposition of the path
into individual worldlines is unique (which actually simplifies things).
Here we simply illustrate the updates. Since in all cases we know how to compute
the path contribution to the statistics of $Z$ or $G$, graphical representations can always be
converted to precise mathematical expressions for the acceptance ratios which account for
the ratio of the configuration weights and probabilities/probability densities of applying
a particular type of update.

Figures \ref{PS:fig:cont_space} and \ref{PS:fig:swap} show an ergodic set of updates which would allow one
to efficiently simulate continuous space systems. The \textit{Open/Close} and
\textit{Insert/Remove} pairs of updates are reminiscent of those in lattice systems. The
\textit{Draw/Erase} pair naturally combines in one update both space and time motion of the
end-point and is essentially a literal implementation of the draw-and-erase algorithm. The
\textit{Swap} update is equivalent to the version of the \textit{Jump} update which involves
reconnection of the particle worldlines. In \textit{Swap} all modifications of the path are restricted
to occur between the two dashed lines, see Fig.~\ref{PS:fig:swap}

The rules for collecting statistics to the diagonal and
off-diagonal properties are similar to lattice models.

\begin{figure}
\resizebox{\hsize}{!}{
\centerline{\includegraphics[width=200pt, height=145pt] {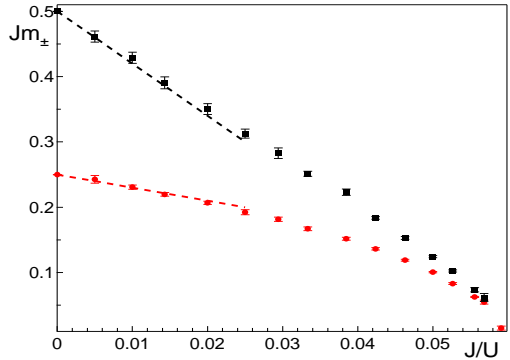}}
}
\caption{Effective mass for
particle (circles) and hole (squares) excitations in the 2D Bose Hubbard model at unity filling as a function of
the hopping-to-interaction ratio $J/U$. The exact results at $J/U=0$ are $m_{+} =0.25/J$ and $m_{-}
=0.5/J$. The dashed lines  show the lowest-order in $J/U$
corrections to the effective masses. Close to the critical point
the two curves overlap, revealing the emergence of
the particle-hole symmetry implied by criticality-induced emergent Lorentz invariance. The sound velocity 
of the relativistic spectrum in the Lorentz-invariant regime is
found to be $c/J=4.8\pm 0.2$.  (Reproduced from Ref.~\cite{PS:BH_2D}.)} \label{PS:fig:eff_mass}
\end{figure}

\subsection{Momentum Conservation in Feynman  Diagrams}
\label{PS:subsec:diagrams}

Diagrammatic Monte Carlo (see, e.g., Ref.~\cite{PS:vH_K_P_S} and references therein) is a technique of sampling entities expressed in terms of  Feynman's diagrammatic series by
a Markov process. The configurational space of the process consists of Feynman's diagrams with fixed internal variables.---It is the sampling process itself that 
accounts for the integration over these variables, on equal footing with summation over the order and topology
of the diagrams. In the process of producing  the Markov chain of diagrams, generating the $(n+1)$-st diagram is performed by applying one of a few elementary updates to the $n$-th diagram. 
The elementary updates  are supposed to change the structure of a diagram (by adding/removing/reconnecting a small number of propagators) and the values of (a small number of) internal variables.
In the space-time representation, updating internal variables causes no problem since these are nothing but the space-time points corresponding to the ends of
propagators, their  values being naturally generated with appearance of new propagators, and abandoned when the corresponding propagators are removed.
In momentum and/or frequency representation, the situation is quite different.  The momentum (for briefness, from now on we speak of momentum only) of a given propagator 
is not independent of the momenta of other propagators in view of the momentum conservation constraint taking place at each vertex. Adding/removing a propagator to/from a diagram
would thus involve a change of momenta of other propagators, rendering  the updating routine complicated and less efficient. 

A simple and efficient way out is provided by the ``worm" idea of working in an extended configurational space. The additional class of diagrams that we need consists of  diagrams featuring
worms, by which here we mean vertices with non-concerving momenta. Clearly,  the minimal non-trivial number of worms is two. Normally, working with no more than two worms proves sufficient.
Note that if the algebraic sum of all the momenta at one of the two worms is $\vec{\delta}$, then its counterpart at the other worm is $-\vec{\delta}$, so that $\vec{\delta}$ is the only continuous parameter
associated with the pair of worms.

A crucial observation now is that if all the structural updates of diagrams are performed in the subspace of diagrams with worms and the ends of the propagator(s) to be added/removed/reconnected
are linked to the two worms, then the worms will readily ``absorb" the residual momentum, $\vec{k}_{\rm res}$, associated with the update,  the only consequence  for the
worms being $\vec{\delta}\to \vec{\delta}+\vec{k}_{\rm res}$. Details on implementing this idea can be found in Ref.~\cite{PS:vH_K_P_S}. The overall updating scenario is as follows. Switching between physical and worm sectors of the configurational space is achieved by a pair of complementary updates creating/deleting a pair of worms at the ends of a propagator, with simultaneously changing the momentum of this propagator. The rest of the updates is performed in the worm sector. The efficiency of the scheme is achieved by introducing an update that  translates worms along one of the propagators attached to the worm vertex. In the updated diagram, the conservation of momentum at the worm's original position
is ensured by appropriately changing the momentum of the propagator along which the translation is being performed. Translating worms along propagators allows one to efficiently sample all their
 positions  and thus apply the updates changing the structure of the diagram---associated with the worms, as discussed above---in a generic way.

\section{Illustrative Applications}

\subsection{Optical-Lattice Bosonic Systems}
\label{PS:subsec:OL}

Bosons in optical lattices, being accurately described by the Hubbard Hamiltonian (\ref{PS:BHmodel}) \cite{PS:Jaksch}, are perfectly suitable for simulations by worm algorithm.
With a standard desktop computer, one can simulate equilibrium properties of 3D systems with  $200^3$ lattice sites. In Refs.~\cite{PS:BH_3D,PS:BH_2D} it has been demonstrated that this approach allows one to obtain precision results  for  equations of state and produce an accurate phase diagram of the system. It is also possible to trace  the evolution of the particle/hole spectrum of elementary excitations with decreasing interaction strength, from the strong coupling limit down to the critical point of the Mott-insulator--to--superfluid quantum phase transition. Especially interesting is the vicinity of the quantum critical point, where the emergent Lorentz invariance brings about particle-hole symmetry, see Fig.~\ref{PS:fig:eff_mass}.
Along with the pure single-component bosonic Hubbard model, one can simulate multi-component and disordered systems, see Figs.~\ref{PS:fig:phase_diag_2cmp}-\ref{PS:fig:disord_bos}.

The lattice bosonic systems are in the focus of Optical
Lattice Emulator project supported by DARPA and
aimed at the development, within the next few years,
of experimental tools of accurately mapping phase diagrams
of lattice systems by emulating them with ultracold
atoms in optical lattices. The numerically exact solutions for real experimental systems will be used for validating the emulators.
The first successful validation of the emulator of the Bose Hubbard model was reported in Ref.~\cite{PS:validation}. At the heart of the  protocol is the direct comparison of the
experimental time-of-flight images with the theoretical ones. The latter are produced by time-evolving the single-particle density matrix obtained in a direct simulation of 
a given number of atoms in a trap, see Fig.~\ref{PS:fig:TOFseries}.

\begin{figure}
\resizebox{\hsize}{!}{
\centerline{\includegraphics[width=300pt, height=250pt]{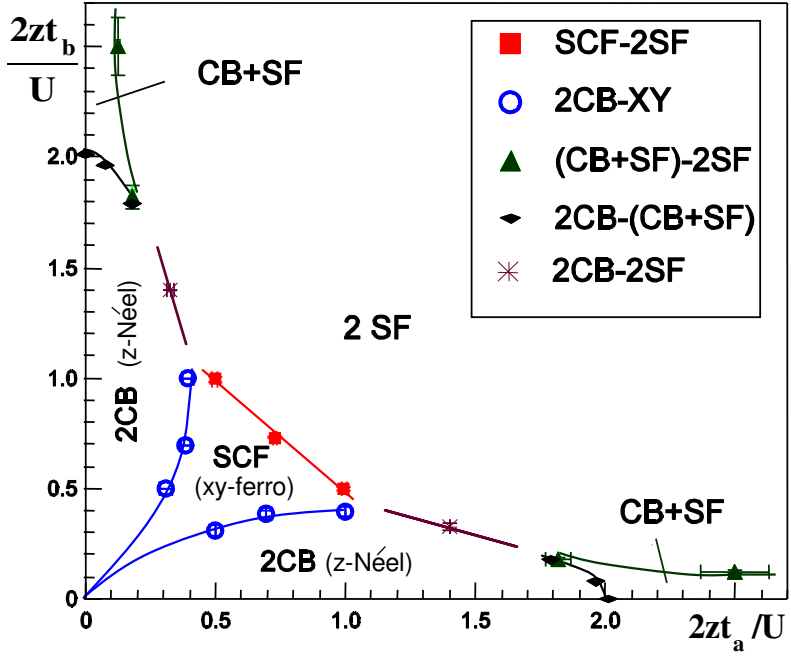}} 
}
\vspace*{1.5cm}
\caption{Groundstate phase diagram of 2D two-component bosonic Hubbard model  at half-integer
filling factor for each component. The on-site interactions within each component are infinitely strong (hard-core limit),
while the inter-component interaction, $U$, is finite.  The hopping elements of the two components are $t_a$ and $t_b$,
the parameter $z=4$ is the coordination number. The revealed phases are as follows. (i) checkerboard solid in both components, a.k.a. $z$-N\'eel phase (2CB),
(ii) checkerboard solid in one component and superfluid in its counterpart (CB+SF), (iii) superfluid in both components (2SF), (iv) super-counter-fluid, a.k.a.
$XY$-ferromagnet (SCF).
The observed transition lines are:
2CB-SCF (first-order),
SCF-2SF (second-order),
2CB-2SF (first-order),
2CB-CB+SF (second-order), and
CB+SF-2SF (first-order). Lines are used to guide an eye. (Reproduced from Ref.~\cite{PS:Soyler}.)}
\label{PS:fig:phase_diag_2cmp}
\end{figure}

\begin{figure}
\resizebox{\hsize}{!}{
\centerline{\includegraphics[angle=0,scale=0.3, angle=-90]{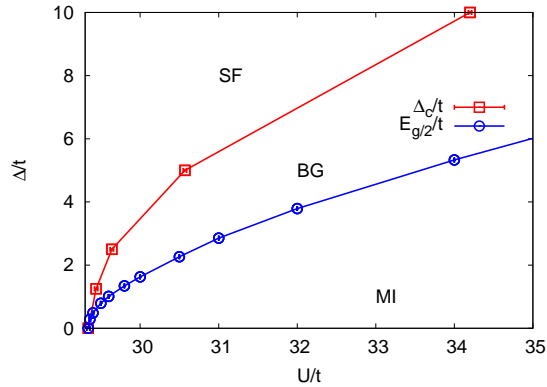}}
}
\caption{Groundstate phase diagram of the disordered 3D Bose-Hubbard model (at unity filling)  in the vicinity of the point of the superfluid(SF)--to--Mott-insulator(MI) quantum phase transition.
The $E_{g/2}(U)$ curve marks the
Bose-glass(BG)--Mott-insulator transition boundary according to the conjecture that the transition occurs when the bound of disorder reaches the half-gap, $E_{g/2}$, of the pure Mott insulator.
Error bars are shown, but are smaller than point sizes. (Reproduced from Ref.~\cite{PS:disorder_3D}.)}
\label{PS:fig:disord_bos}
\end{figure}

\begin{figure}
\resizebox{\hsize}{!}{
\centerline{\includegraphics[angle=0,scale=1.2]{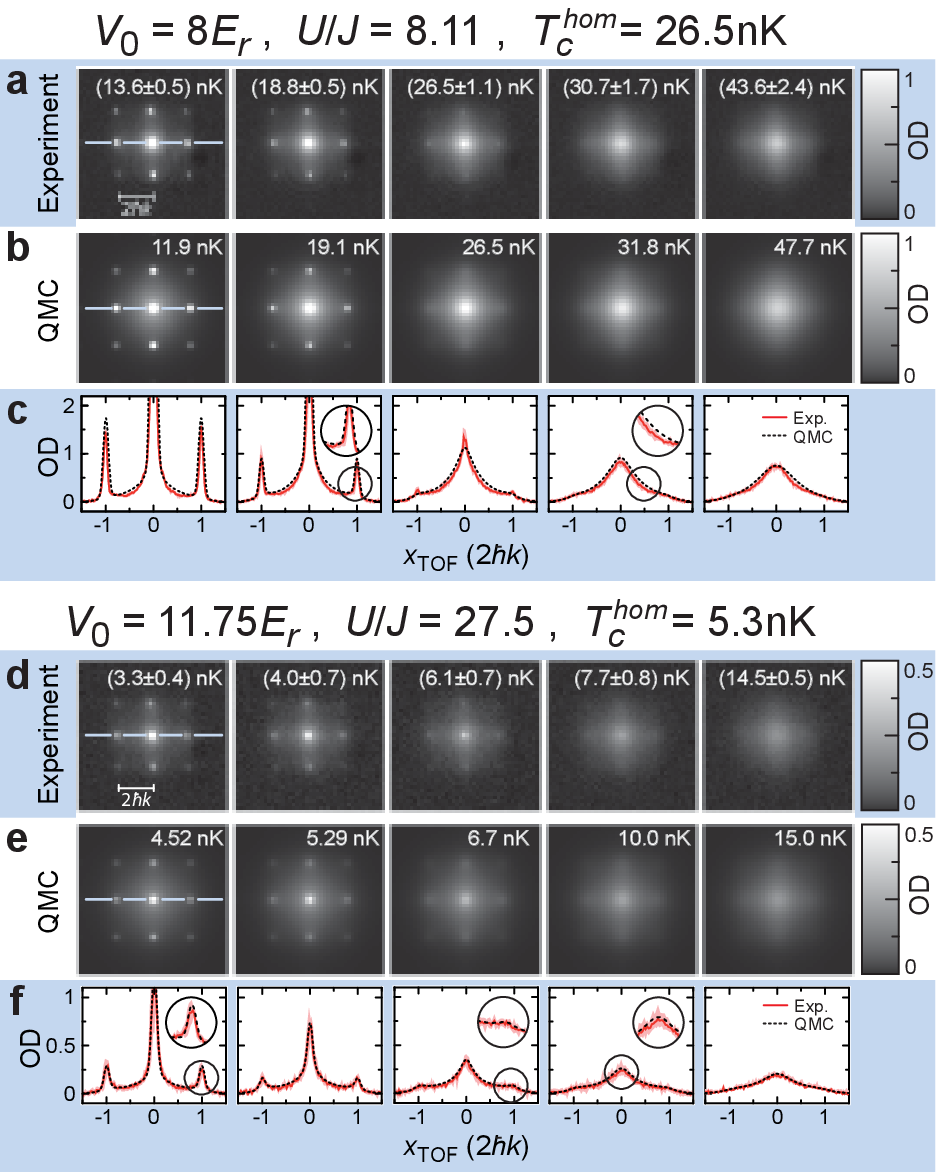}}
}
\vspace*{0.5cm}
\caption{Comparison of experimental and simulated time-of-flight distributions: Shown the integrated column density $n_{\perp}(x,y)$ represented by the optical density  as obtained from the experiment and the QMC simulations for different temperatures and two lattice depths.  (Reproduced from Ref.~\cite{PS:validation}; see this reference for more detail.)}\label{PS:fig:TOFseries}
\end{figure}

\subsection{Supersolidity of Helium-4}
\label{PS:subsec:supersolid}

A supersolid is a quantum solid that can support a dissipationless flow of its own atoms. Here the term `solid' is understood 
in the most general context of any (regular or amorphous, continuous-space or lattice) state with broken translation symmetry.

The modern age of supersolidity of bosonic  crystals in continuous space began with the discovery by Kim and Chan of non-classical rotational inertia (NCRI)
in solid $^4$He \cite{PS:KC1,PS:KC2};  for reviews of further developments  in the field, including direct observation of a superflow, as well as preceding work, see Refs.~\cite{PS:Prokofiev_review,PS:Balibar_review,PS:Svistunov} and references therein. 
In the combined experimental and theoretical effort aimed at understanding the microscopic picture behind the effect, the first-principles simulations of regular and disordered solid
$^4$He play a very important part. Here we present some numeric results shedding a direct light on microscopic mechanisms of supersolidity in $^4$He.

\begin{figure}
\resizebox{\hsize}{!}{
\centerline{\includegraphics[width=160pt, height=200pt, angle=-90]{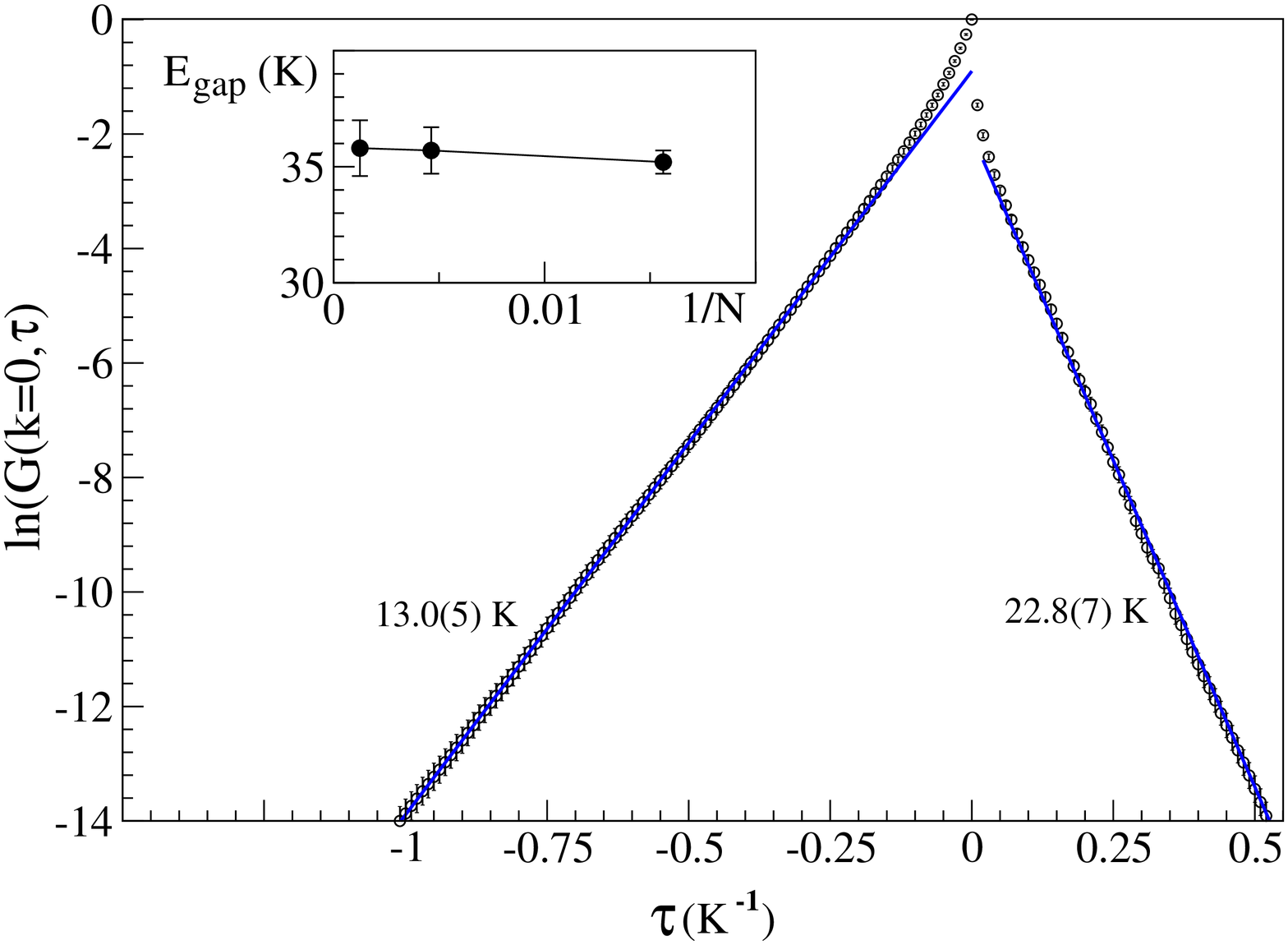}}
}
\vspace*{-0.5cm}
\caption{Single-particle Green's function $G({\bf k}=0,\tau)$
computed by worm algorithm for hcp $^4$He at the melting density $n_{\circ}$=0.0287 $\,$\AA$^{-3}$~
and $T=0.2$~K. Symbols refer to numerical data,
solid lines are fits to the long-time exponential decay.
The given numerical values are the interstitial ($\Delta_{\rm I}=22.8 \pm 0.7$ K)
and the vacancy ($\Delta_{\rm V}=13.0 \pm 0.5$ K) activation energies,
inferred from the slopes of $G$.
The straight-line asymptotic behavior indicates that  finite-temperature corrections are negligible.  The inset shows the vacancy-interstitial gap $E_{\rm gap} = \Delta_{\rm I}+\Delta_{\rm V}$
 for different system sizes, proving that the results have reached their macroscopic limit. (Reproduced from Ref.~\cite{PS:vacancy}.)}
\label{PS:fig:green}
\end{figure}

\begin{figure}
\resizebox{\hsize}{!}{
\centerline{\includegraphics[width=160pt, height=220pt, angle=-90]{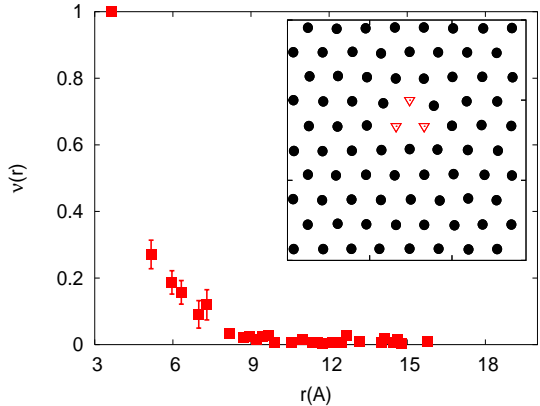}}
}
\caption{The vacancy-vacancy correlation function $\nu(r)$ as a function of the distance $r$ between 
the vacancies shows that three vacancies easily cluster and form a tight bound state. 
The inset shows a typical snapshot of a layer of atomic positions averaged over the time interval 
$[0, \beta]$(filled black circles). It is seen  that the three vacancies (triangles) have a tendency to cluster in layers. (Reproduced from Ref.~\cite{PS:vacancy}.) }
\label{PS:fig:vv}
\end{figure}

\begin{figure}
\resizebox{\hsize}{!}{

\centerline{\includegraphics[width=290pt, height=420pt, angle=-90]{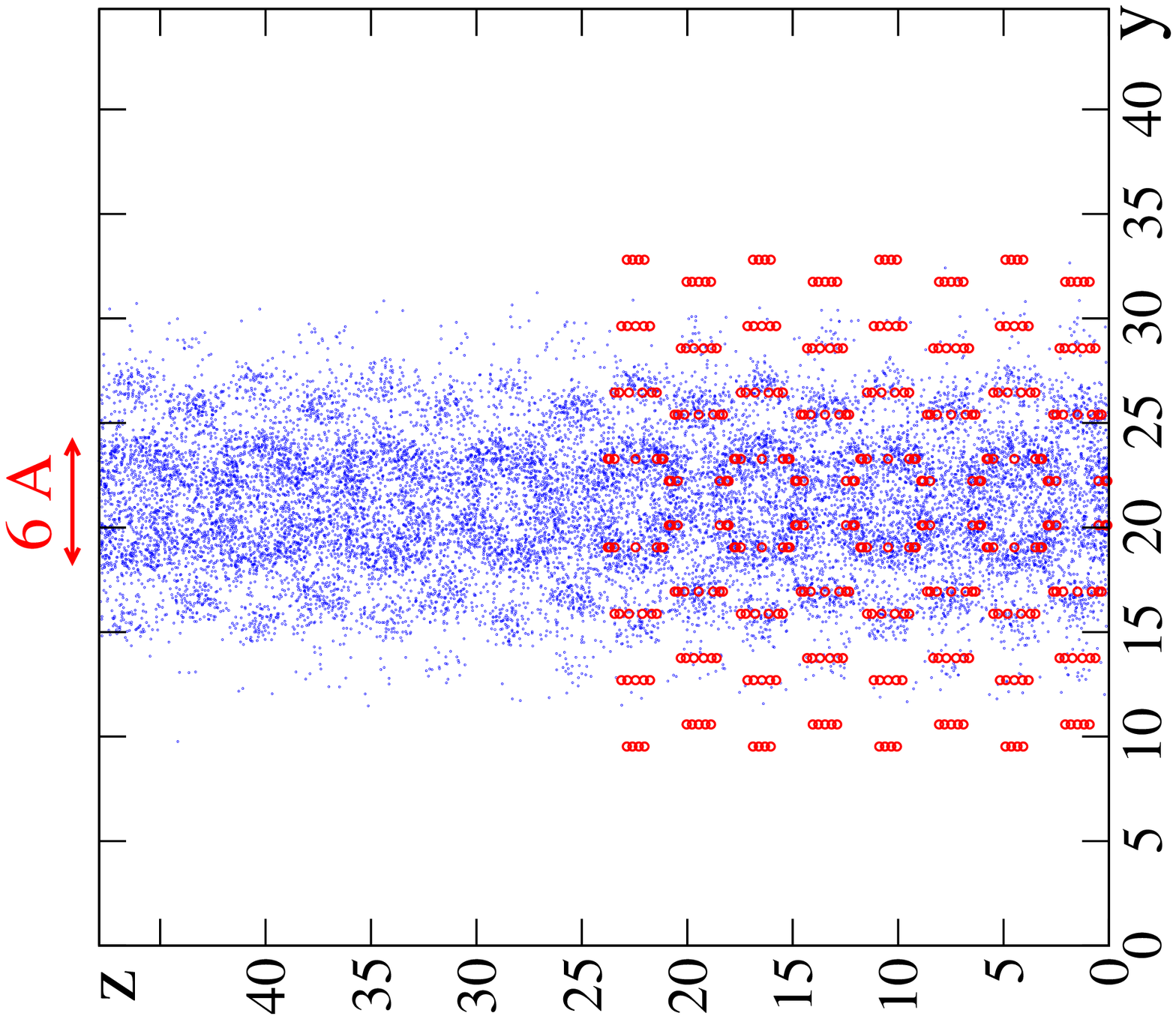}}
}
\vspace*{1.0cm}
\caption{Luttinger liquid in the core of the screw dislocation in solid $^4$He revealed by columnar winding-cycle map. Simulations correspond to  
the temperature $0.25\,$K and density 0.0287$\,$\AA$^{-3}$.  View is along the $x$-axis in the basal plane---perpendicular to
the core. Shown with large dots (in the lower half of the plot only) are the atomic positions in initial configuration. 
The unit of length is 1\AA. (Reproduced from Ref.~\cite{PS:screw}.)}
\label{PS:fig:screw}
\end{figure}

To introduce a general theoretical background for interpreting numeric results, we start with a number of rigorous statements. From the field-theoretical perspective, a superfluid/supersolid groundstate---as opposed to an insulating one---is almost trivial, since the  topological constant of motion responsible for the superfluidity is naturally introduced in terms of the phase of the classical matter field, so that phenomenon of superfluidity in a quantum field is simply inherited from the classical counterpart of the latter. An insulating  groundstate is possible only in a quantum field. It is an essentially non-perterbative and thus non-trivial phenomenon; the fact of its existence in bosonic systems has been recently denied by P.W. Anderson \cite{PS:Anderson}. 

A proof of existence of insulating groundstates  in bosonic crystals
immediately follows from the theorem \cite{PS:PS2005} stating that a necessary condition for supersolidity is the presence of either vacancies, or interstitials, or both. With this theorem, one just needs to make sure
that there are  groundstates where  creating a vacancy and an interstitial cost finite energy. The latter is known to be the fact at least since Andreev and Lifshitz analysis \cite{PS:AL}: The energy for creating a vacancy/interstitial is positive, and arbitrarily close to the classical-crystal value, in the limit of strong interaction/large particle mass.

The theorem of Ref.~\cite{PS:PS2005} offers a reliable protocol of numerical proof that the groundstate of  perfect $^4$He hcp crystal is an insulator. It is sufficient to demonstrate that (i) creating 
a vacancy/interstitial in the simulation box costs a finite energy, the value of which does not vanish with increasing the system size and decreasing the temperature, (ii) a state with finite concentration of vacancies/interstitials---that could potentially differ from the single-vacancy/interstitial situation due to collective effects---is unstable in the thermodynamical limit with respect to aggregation 
(i.e. a crystal at $T=0$ purges itself from the vacancies and interstitials). Both facts are demonstrated in Figs.~\ref{PS:fig:green} and \ref{PS:fig:vv}. (The aggregation of interstitials has not been studied since
these cost much more energy than the vacancy, rendering the scenario of interstitial-induced supersolitity not realistic.)

Having established that the perfect hcp $^4$He crystal is not  supersolid, one has to explore  {\it disordered} scenarios, when the superflow in a crystal
is supported by defects. First-principles simulations performed by UMass-ETH-UAlberta-CUNY 
collaboration (briefly reviewed in Ref.~\cite{PS:Svistunov}) have revealed a number of disorder induced mechanisms of supersolidity in $^4$He: superfluid dislocations, 
grain boundaries, ridges, and also a metastable amorphous supersolid, the so-called superglass. Here we confine ourselves with presenting the results for the superfluidity in the
core of a screw dislocation---arguably the cleanest Luttinger liquid system in Nature. 

To visualize spatially inhomogeneous superfluidity in a worm algorithm simulation, one can employ two similar  approaches. 
 One approach is to plot the condensate density map. The other and more general approach (that also
works for lower-dimensional systems with the genuine long-range
order destroyed by fluctuations of phase) is to visualize the
macroscopic worldline loops responsible for non-zero winding
numbers, and thus for the superfluid response. Identifying these
loops in a given worldline configuration, projecting them from the
$(d+1)$ dimensions onto a plane in the real space, and performing
the average over a representative set of configurations, one
obtains the {\it winding-circle map} of the superfluid region. The Luttinger liquid core of
a screw dislocation visualized with this technique is shown in Fig.~\ref{PS:fig:screw}.

\subsection{Problem of Deconfined Criticality. Flowgram Method}
\label{PS:subsec:DCP}

The standard Ginzburg-Landau-Wilson (GLW)  scenario of critical
phenomena excludes generic  continuous transitions between states which break different
symmetries, thus implying that the transition, if exists, is of the first order.
An intriguing possibility of breaking GLW paradigm was proposed in Refs.~\cite{PS:Motrunich,PS:Senthil1,PS:Senthil2}
for the so-called {\it deconfined critical points} (DCP) in two spatial dimensions. Nowadays the problem of DCP is one of the most exciting, and yet
controversial topics in the theory of phase transitions. Remarkably, the field-theoretical model for deconfined criticality---to be referred below 
as the DCP action---is given by  two identical complex-valued  classical fields coupled to a gauge vector field, in three dimensions.
Despite its simplicity and apparent closeness to the single-component counterpart (known to be in the inverted XY universality class
of continuous phase transitions), the DCP action is not amenable to reliable analytic treatments because of its runaway renormalization flow to strong coupling at large scales.
To establish the order of the phase transition in this model one has to resort to numerics.

\begin{figure}
\resizebox{\hsize}{!}{
\centerline{\includegraphics[width=280pt, height=220pt]{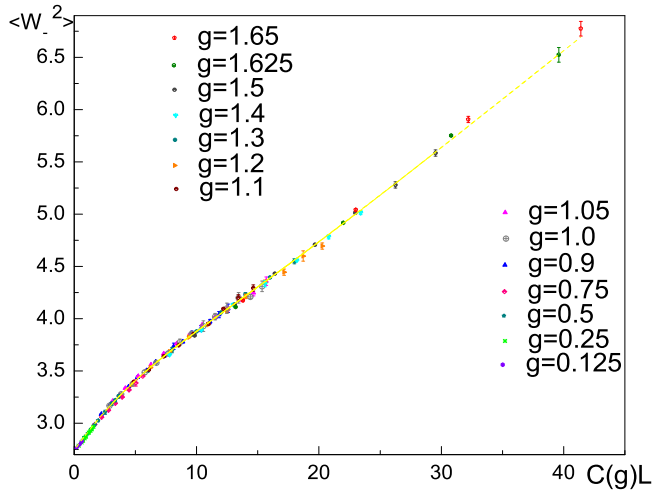}}
}
\vspace*{-1cm}
\caption{Data collapse for the flows of the SU(2) symmetric DCP action.  
The  line is a fit representing the master curve. The horizontal
axis is the scale reduced variable $C(g)L$.  (Reproduced from Ref.~\cite{PS:DCP2}.)}
\label{PS:fig:flowgram_collapse}
\end{figure}

\begin{figure}
\resizebox{\hsize}{!}{
\centerline{\includegraphics[width=280pt, height=210pt]{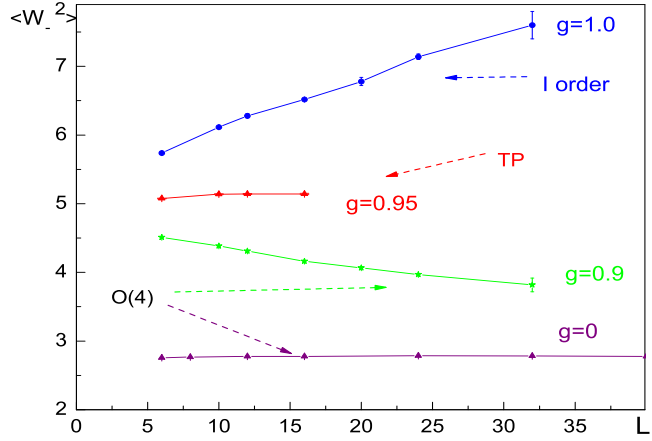}}
}
\vspace*{-1cm}
\caption{ Flowgrams for the short-range
model $V_{ij} =  g\delta_{ij}$. The lower horizontal line features the O(4) universality scaling behavior; for $g<g_c \approx 0.95$ all flows are attracted to this
line. The upper horizontal line is the tricritical separatrix (marked as TP).
Above it, flows diverge due to the first-order transition
detected by the bi-modal distribution of energy. (Reproduced from Ref.~\cite{PS:DCP2}.) }
\label{PS:fig:flowgram_short}
\end{figure}

In Refs.~\cite{PS:DCP1,PS:DCP2}, the order of the phase transition in the DCP action was studied by the worm algorithm (for the U(1)$\times$U(1)- and SU(2)-symmetric actions, respectively).
Within the given universality class, the optimal choice of  microscopic model is the high-temperature expansion (cf. Sec.~\ref{PS:subsec:HTE}) of a (cubic-)lattice theory, 
resulting in the bond-current model with the following partition function (for the sake of definiteness, we present the answer for the SU(2)-symmetric case).
\begin{equation}
Z=\sum_{\{
J \} } {\cal Q}_{\rm site}\: {\cal Q}_{\rm bond} \: \exp(-H_J), ~~~~H_J=(g/2)\!\!\!\!\!\!\!  \sum_{ i,j;\, a,b;\, \mu=1,2,3} \!\!\!\!\!  
I^{(a)}_{i,\mu }\, V_{ij} \, I^{(b)}_{j,\mu },~
\label{PS:Z_DCP}
\end{equation}
\begin{eqnarray}
{\cal Q}_{\rm site}=\prod_i  \frac{{\cal N}^{(1)}_i! \, {\cal
N}^{(2)}_i!}{(1+{\cal N}^{(1)}_i+ {\cal N}^{(2)}_i)!},\quad
 {\cal N}^{(a)}_i=\frac{1}{2} \sum_{\mu} J^{(a)}_{i,\mu}, ~~~ {\cal Q}_{\rm bond} = \prod_{i,a, \mu } \frac{t^{J^{(a)}_{i,\mu}}}{J^{(a)}_{i,\mu }!} .
\nonumber
\end{eqnarray}
Here $J^{(a)}_{i,\mu }$ is an integer non-negative bond current of the component $a=1,2$, living on the bond $(i,\mu)$.
The bond subscript of a current is represented by the site, $i$, and direction $\mu=\pm 1,\pm 2,\pm 3$ from this site. 
The direction-dependent subscript reflects the fact  that for a given geometric
bond there are {\it two} different bond currents of the same component $a$, and $I^{(a)}_{i,\mu } = J^{(a)}_{i,\mu }-J^{(a)}_{i+\hat{\mu},
-\mu }$ is their algebraic sum (here $\hat{\mu}$ is a unit translation vector in the direction $\mu$).
The bond currents are subject to the conservation constraint on each site: 
$ \sum_{\mu } I^{(a)}_{i,\mu}=0$.
The parameter $t$ controls the strength of the lattice gradient term for the complex fields in the DCP action, 
and $g$ is the coupling constant for the interaction between the complex and gauge fields. The integration over the gauge field results in the long-range interaction, $V_{ij}$, 
between the currents. The Fourier transform of $V_{ij}$
is given by $V_{\bf q} = 1/\sum_{\mu=1,2,3 } \sin^2 (q_{\mu}/2)$
and implies an asymptotic behavior $V_{ij} \sim 1/r_{ij}$ at large
distances. It is this Coulomb asymptotic tail of the current-current interaction that leads to a qualitative difference of the DCP action from its short-range counterparts. 

With its closed-loop structure enforced by the current conservation constraint  and the positive-definitness of the weighting factors,  the model (\ref{PS:Z_DCP})
is in the domain of applicability of the worm algorithm, each of the two components being updated by its individual pair of worms. The results of simulations
of both U(1)$\times$U(1)- and SU(2)-symmetric actions lead to an unfortunate for the DCP theory conclusion that the phase transition is of the first order. 
The definitive conclusion is based on the {\it flowgram method} \cite{PS:DCP1}
of finite-size analysis.  The key idea is to demonstrate that the universal large-scale
behavior at $g\to 0$ is identical to that at some finite coupling $g=g_{\rm coll}$
where the nature of the transition can be easily revealed. The procedure is:
\begin{enumerate}
\item[(i)] Introduce a definition of the critical point for a finite-size system
of linear size $L$ consistent with the thermodynamic limit and insensitive
to the order of the transition. 
Specifically,  for any given $g$ and $L$ the critical value of  $t$ was defined by the requirement that the ratio of statistical weights of configurations with and without windings 
be equal to a fixed constant.
\
\item[(ii)] At the critical point, calculate a quantity $R(L,g)$ that is
supposed to be scale-invariant for a continuous phase transition
in question, vanish in one of the phases and diverge in the other. 
Specifically, once can take $R(L,g)$ to be the variance of the winding number in the counter-flow channel:  $R(L,g)\equiv \langle W_-^2 \rangle\equiv \sum_\mu \langle (W_{1,\mu}-W_{2,\mu})^2\rangle$,
where $W_{a,\mu}$ is the winding number of the component $a$ in the direction $\mu$. (See Ref.~\cite{PS:DCP1} for the motivation of this choice.)
\
\item[(iii)] Perform a data collapse for flowgrams of $R(L,g)$,
by rescaling the linear system size, $L \to C(g)L$, where $C(g)$ is
a smooth and monotonically increasing function of the
coupling constant $g$. [In the present case, it is  {\it a priori} known that  $C(g\to 0) \propto g$.]
\end{enumerate}

A collapse of the rescaled flows within an interval $g\in [0,\, g_{\rm coll}]$
implies that the type of the transition within the interval remains the same,
and thus can be inferred by dealing with the $g=g_{\rm coll}$ point only. 
Since the  $g\to 0$ limit implies large spatial scales, and, therefore, model-independent
runaway renormalization flow pattern, the conclusions are universal.

As is seen in Fig.~\ref{PS:fig:flowgram_collapse}, the flows for the DCP action collapse perfectly  in the region 
$0.125\leq g <g_{\rm coll}=1.65$.  The rescaling function $C(g)$ exhibits the expected linear behavior $C(g)\propto
g$ at small $g$.

In accordance with the above-outlined logic of flowgram method, the flow collapse within the interval $g\in [0, g_{\rm coll}]$ 
proves that the order of the transition within this
interval does not change.  The mere fact of the data collapse on a master curve with a finite slope is not sufficient to conclude that
the transition is of the first order.
What appears to be a (characteristic of the first-order transition) diverging behavior in Fig.~\ref{PS:fig:flowgram_collapse} might be just a reconstruction
of the flow from the O(4)-universality (at $g=0$) to a novel
DCP-universality at strong coupling. To complete the proof,
one has to determine the nature of the transition for $g=g_{\rm coll}$. 
In this parameter range, the standard technique of detecting discontinuous transitions
by the bi-modal energy distribution becomes feasible. At $g=1.65$ it becomes possible to clearly see  a bi-modal energy distribution that gets more and more pronounced 
with increasing the systems size.  
This leads to a  conclusion that the whole phase transition line for small
$g$ features a generic weak first-order transition.  Driven by long-range interactions, this behavior develops on
length scales $\propto 1/g \to \infty$ for small $g$ and thus is universal.

Finally, it is very instructive to contrast the flowgram  for the DCP action (\ref{PS:Z_DCP}) with the flowgram for the short-range counterpart of  (\ref{PS:Z_DCP}) , where $V_{ij} =  g\delta_{ij}$.
The flows for the short-range model are presented in Fig.~\ref{PS:fig:flowgram_short}. As opposed to the DCP action case, it is impossible to collapse the data on a single master curve by
re-scaling $L$. Now the flows clearly reveal a tri-critical point separating the second-order part of the phase transition line from the first-order part.

\section{Conclusions and Outlook}
Worm Algorithm is a technique for performing efficient updates of 
configurations that have the form of closed paths/loops. 
The closed-loop structure imposes topological constraints and thus creates
ergodicity problems for local updates in large system sizes. 
Worm Algorithm works in an extended configuration space containing 
all the original configurations as well as configurations with open loops. 
All updates are local and are performed exclusively at the open loop ends,
referred to as worms. In most cases, the open-loop
configurations themselves are of prime physical interest being
associated with correlation functions, such as the single-particle
Green function. It was demonstrated for a variety of systems and
universality classes that WA eliminates problems with ergodicity
and critical slowing down, successfully competing even with
model-specific cluster algorithms \cite{PS:Sokal2007}.
At the same time, it is a flexible approach with a broad range of applications in
many-particle bosonic and spin systems. It readily produces loops winding around the system,
allows efficient simulations of off-diagonal correlations, grand canonical ensembles,
disordered systems, etc. At the moment WA has no competitors among unbiased first-principles
approaches for bosons with strong interactions between the particles,
as well as with strong external---both regular and disordered---potential. 
It is easy to predict that in the nearest future WA will remain the method of choice 
for detailed studies of non-trivial strongly correlated bosonic systems 
(multicomponent, disordered, with long-range interactions, solid and supersolid $^4$He, etc.)
Recently, WA has proved indispensable for guiding experimental efforts 
in creating optical lattice emulators; it will continue playing this important role.

In its most general form, the idea of WA is to work in an enlarged configuration space
which includes configurations violating constraints characteristic of the physical
configurations. Nowadays worm-type updates and/or worm-type estimators
for the Green function are an integral part of
many other state-of-the-art lattice Monte Carlo algorithms
\cite{PS:Wiese,PS:DT,PS:Sandviklast,PS:Kawashima}. It is only upon the implementation of the worm-type
updates it became possible to overcome the critical slowing down in the Stochastic Series Expansion
scheme \cite{PS:Sandviklast}.  Important improvements have been made to achieve maximal
efficiency of the worm-type updates by suppressing bouncing: The so-called directed (guided) loop and geometrical
WA \cite{PS:Sandviklast,PS:Alet}, appear to be an optimal combination in
terms of universality and performance. Successful applications of WA 
in high-energy physics \cite{PS:WA_HEP_1, PS:WA_HEP_2, PS:WA_HEP_3}, with exciting
most recent developments  \cite{PS:Wenger,PS:Wolff,PS:Forcrand}, render the approach 
interdisciplinary. Whenever a new representation for Quantum Monte Carlo appears, 
the generalized WA idea may prove useful
for developing an efficient updating strategy.

\textit{Acknowledgments --} We acknowledge  support from the NSF (Grant No. PHY-0653183) and from the Army Research Office with funding from the DARPA OLE program, and the hospitality of the Aspen Center for Physics where this chapter was written.




\end{document}